\documentclass[pre,twocolumn,amssymb,superscriptaddress]{revtex4-1}
\usepackage{amsmath,amsfonts,graphics,float}
\usepackage[caption=false]{subfig}
\usepackage{graphicx}
\usepackage{epsfig}
\usepackage{latexsym}
\usepackage{bm}
\usepackage{mathtools}

\newcommand{\be}{\begin{equation}}
\newcommand{\ee}{\end{equation}}
\newcommand{\bea}{\begin{eqnarray}}
\newcommand{\eea}{\end{eqnarray}}
\begin{document}

\title{Kinetically constrained model for gravity-driven granular flow and clogging}

\author{Gregory Bolshak}
\email{gregory7@mail.tau.ac.il}
\affiliation{School of Mechanical Engineering and Sackler Center for Computational Molecular and Materials Science, Tel Aviv University, Tel Aviv 69978, Israel}
\author{Rakesh Chatterjee}
\affiliation{School of Mechanical Engineering and Sackler Center for Computational Molecular and Materials Science, Tel Aviv University, Tel Aviv 69978, Israel}
\author{Rotem Lieberman}
\affiliation{Rafael Advanced Defense Systems, Haifa 31021, Israel}
\author{Yair Shokef}
\email{shokef@tau.ac.il}
\affiliation{School of Mechanical Engineering and Sackler Center for Computational Molecular and Materials Science, Tel Aviv University, Tel Aviv 69978, Israel}

\begin{abstract}
We add extreme driving to the Kob-Andersen kinetically-constrained lattice-gas model in order to mimic the effect of gravity on dense granular systems. For low particle densities, the current that develops in the system agrees at arbitrary field intensity with a mean field theory. At intermediate densities, spatial correlations give rise to non-monotonic dependence of the current on field intensity. At higher densities, the current ultimately vanishes at a finite, field-dependent jamming density. We supplement the study of this bulk behavior with an investigation of the current through a narrow hole. There, lateral flow decreases the local density in front of the hole. Remarkably, the current through the hole quantitatively agrees with a theoretical prediction based on the bulk current at the measured local density.
\end{abstract}

\date{\today}

\maketitle

\section{Introduction}

Jamming is the transition from a flowing state to a rigid or arrested state as density increases in an assembly of particles~\cite{Liu-Nature, Hecke-JPhysCond}. Colloidal suspensions, glass-forming liquids, and granular materials show heterogeneous dynamics as they approach jamming~\cite{Berthier-PRE2003, Leonard-JCP2010, Pastore, Garrahan-PRL2002, Berthier-JCP2003, Marinari-EPL, Whitelam-PRE, Teomy-PRE2015}. Clogging is when the flow of granular materials through a bottleneck~\cite{clog-bottleneck, Zuriguel_2005, Zuriguel_2011, Zuriguel_2012} or pipe~\cite{clog-pipe} cease to exist. Jamming is a bulk phenomena that occurs throughout the system, while clogging is identified as inability to flow through a specific narrow region in the system where particles get stuck. This eventually causes the flow to stop also in other regions. There have been studies of the crossover between jamming and clogging in heterogeneous environment~\cite{Peter-Srep2018}. Many systems in nature and in industrial applications exhibit jamming or clogging in very complicated geometries~\cite{Exp-1, Exp-3}, and it is important to understand how confinement induces the jamming or clogging of granular matter. Experiments on granular systems~\cite{Abate-PRE} show steady growth in dynamical heterogeneity as the relaxation time increases with increasing density. We employ kinetically-constrained models~\cite{Sellitto_ASEP_KCM_PRL, Shokef-EPL, Turci_Pitard_Driving_KCM, Teomy-PRE2014, Teomy-PRE2017, Mallick2018} to construct a lattice model for describing the physical mechanisms involved in jamming. These models do not rely on any type of interaction between the particles other than excluded volume interaction and are supplemented by specific kinetic rules that control the movement of particles depending on the local density around them. These dynamical restrictions yields glassy behavior in the system.

In this paper, we examine the effect of an external field on jamming in the two-dimensional Kob-Andersen~\cite{Kob-PRE} lattice gas with kinetically-constrained dynamical rules. We implement the external field in a way that, to our understanding, better represents gravity-driven flow of granular materials, compared to previous studies~\cite{Sellitto_ASEP_KCM_PRL, Turci_Pitard_Driving_KCM}. We measure the bulk current as a function of bulk density and observe that the current vanishes at a jamming density which depends on the applied external field. We also measure the bulk current as a function of the external field for the whole range of bulk densities. To understand the microscopic origin for the dynamically jammed states, we find that these states exhibit some rare mobility regions along with other regions of higher mobility due to  local relaxation. Increasing or decreasing local mobility in some region sometimes facilitates the overall cooperative dynamics of the system. Furthermore, we study the effect of confinement on clogging by investigating the behavior of the system when particles are allowed to pass through a narrow orifice. We quantitatively explain the measured current in terms of the current of an effective bulk system with the local density that we measure in front of the orifice.

The paper is structured as follows. In Sec.~\ref{sec:model}, we introduce our model and describe the numerical simulation details. In Sec.~\ref{sec:bulk_current}, we study the bulk current as a function of external field and show the non-monotonic behavior of current. We measure two-point density correlations, focus on the microscopic structure of the system at different parameter regime, and obtain the jamming phase-diagram vs density and field. In Sec.~\ref{sec:orifice}, we study our model with confinement, which mimics the flow of particles through an orifice. Section~\ref{sec:conclusions} summarizes the work.

\section{Model}\label{sec:model}

\begin{figure}[t]
\begin{minipage}[t]{0.45\columnwidth}
\begin{center}
\subfloat[E-Model\label{fig:E-Model}]{
\begin{centering}
\includegraphics[width=0.8\columnwidth]{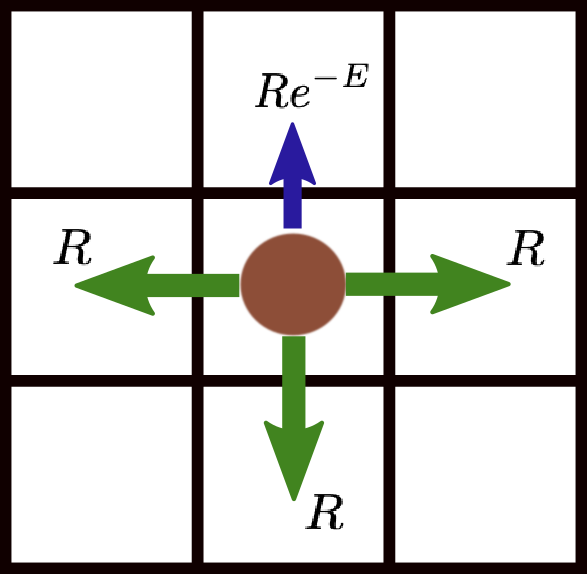}\par
\end{centering}
			}
\end{center}
\end{minipage}\hfill{}
\begin{minipage}[t]{0.45\columnwidth}
\begin{center}
\subfloat[G-Model\label{fig:G-Model}]{
\begin{centering}
\includegraphics[width=0.8\columnwidth]{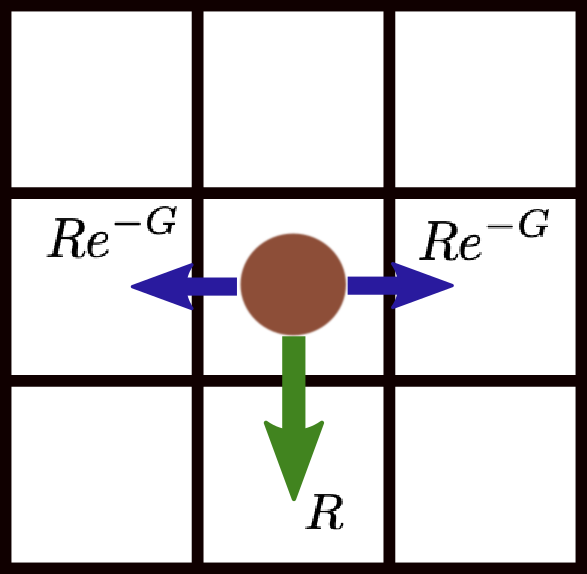}\par
\end{centering}
			}
\end{center}
\end{minipage}	
\caption{(a) E-Model: each particle has a fixed rate to move in three direction denoted by green arrows, and a reduced rate to move against the direction of the field, as shown by the blue arrow. (b) G-Model: each particle has a fixed rate to move in the direction of the field, as shown by the green arrow, and a reduced rate to move in each of the transverse directions, as shown by the blue arrows, with no movement against the direction of the field. In both models, moves succeed only if the kinetic constraints are satisfied.}
\end{figure}

We use the Kob-Anderson kinetically-constrained model~\cite{Anderson-PRL}, in which
particles are randomly distributed on the square lattice with density $\rho$ and they randomly attempt to move in one of four possible directions. According to this model, the move is allowed only if the target site is vacant and if there are at least two vacant nearest neighbors before and after the move. An external field may be applied in some preferred direction in a manner, which we will refer to as the \emph{E-Model}~\cite{Sellitto_ASEP_KCM_PRL, Turci_Pitard_Driving_KCM}. Then, each particle has a rate $R$ to move along the field or in each of the transverse directions, whereas there is a reduced rate $Re^{-E}$ for particles to move against the external field, see Fig.~\ref{fig:E-Model}.

We suggest that in gravity driven granular flow, the rate of motion along the field should be substantially higher than in the transverse directions. Thus we introduce the \emph{G-model} with a constant rate $R$ to move along the field together with transverse movement with rate $Re^{-G}$, and no movement against the field, as shown in Fig.~\ref{fig:G-Model}. Here particles have a constant rate to attempt moving in the direction of the field $G$, and particle current changes by varying the lateral motion, which in turn influences the acceptance rate of moves along the field. Note that the extreme driving limit of the E-model, $E=\infty$, is equivalent to $G=0$ in the G-model. Thus by increasing $G$, we are changing the rate of motion in the transverse direction and thus extending the E-model to even more extreme driving.

\section{Bulk Current}\label{sec:bulk_current}

In order to understand the behavior of the G-model precisely, we study the dynamics of the system in various regimes. Here we begin with the bulk behavior and  consider a system with periodic boundary condition in both directions. In the low-density regime, we expect correlations to be weak and therefore we can predict the current of the particles in the direction of the field. We define the occupancy index $\eta_{i,j}$ for each site $(i,j)$ in the two-dimensional lattice as:
\begin{equation}
\eta_{i,j}=\begin{cases}
1 & \mathrm{the\,site\,is\,occupied}\\
0 & \mathrm{the\,site\,is\,vacant} \nonumber
\end{cases}
\end{equation}

\begin{figure}[t]
\begin{center}
\includegraphics[width=0.4\columnwidth]{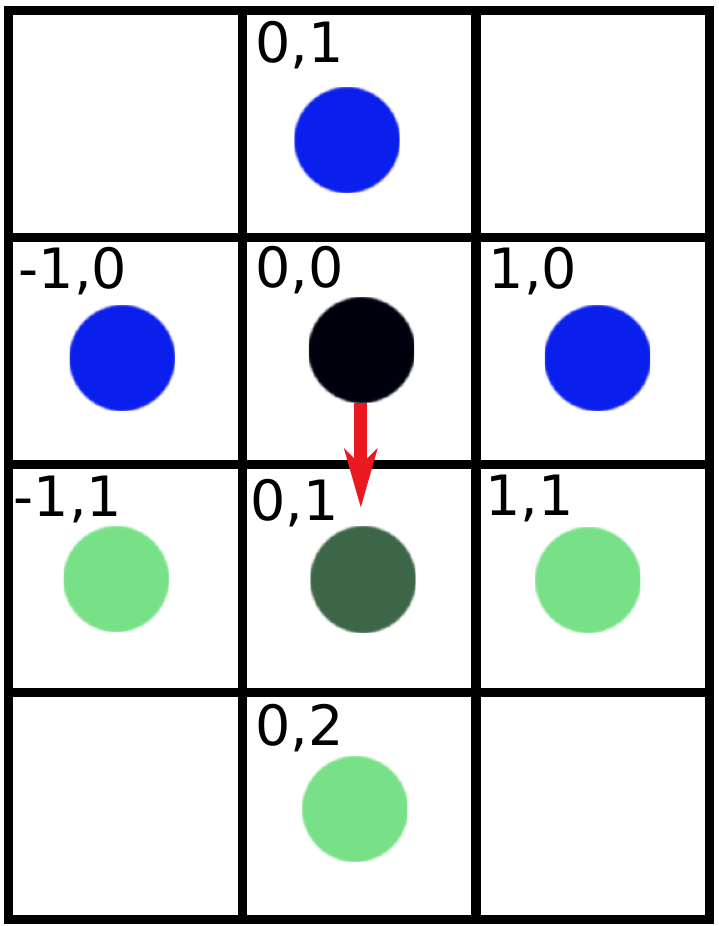}
\caption{When a particle at position (0,0) attempts to move one step in the direction of the field, the target site (0,1) should be vacant, but also at least one of the (blue) sites neighboring the origin site and at least one of the (light green) sites neighboring the target site should be vacant to obey the kinetic constraint.}
\label{fig:occupancy}
\end{center}
\end{figure}

Now we can write the mean current in the direction of the field, based on the kinetic constraints.  It can be clearly seen from Fig.~\ref{fig:occupancy} that when a particle at $(0,0)$ attempts to move to the vacant site $(0,1)$, then at least one of the sites denoted by light blue must be empty along with at least one of the sites denoted by light green that must also be empty to satisfy the kinetic constraint. Therefore the current reads,
\begin{equation}
J=  \langle\eta_{0,0}(1-\eta_{0,1})(1-\eta_{-1,0}\eta_{0,-1}\eta_{1,0})(1-\eta_{-1,1}\eta_{0,2}\eta_{1,1})\rangle .
\label{eq:J_eta}
\end{equation}
Without correlation we obtain the mean-field (MF) current,
\begin{equation}
J_{MF}(\rho)=\rho(1-\rho)(1-\rho^{3})^{2} .
\label{eq:MF}
\end{equation}

We run dynamical Monte Carlo simulations with a rejection-free algorithm, which makes the simulations very efficient, on a periodic lattice of dimension $L\times L$ and
typically use $L = 400.$ We first allow the system to relax for $t = 10^6$ time steps and then start measuring the current after the system has reached the steady state. We compare our numerical result with the MF current, first in the two extreme limits: $G=0$ and $G=\infty$, having only longitudinal motion in the later case.

\begin{figure}[h]
\includegraphics[width=\columnwidth]{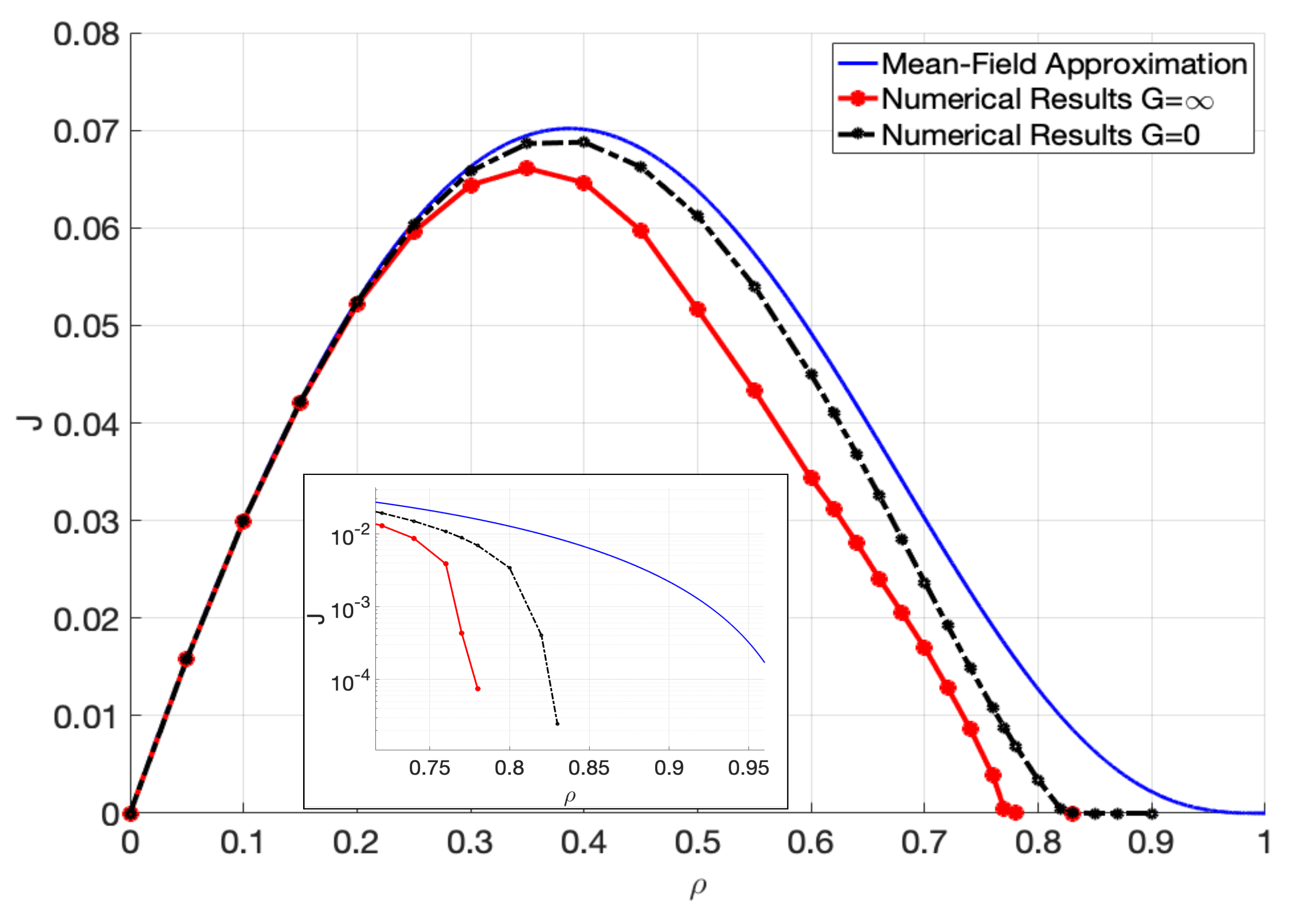}
\caption{Current as a function of density in simulations in two regimes: $G=0$ (black) and $G=\infty$ (red) along with mean-field approximation from Eq.~(\ref{eq:MF}) (blue). They agree at low densities, above which deviations from mean field, and dependence on $G$ start to appear due to increasing correlations. Zoomed plot in at the inset shows precisely the density range where current vanishes.}
\label{MF_graph}
\end{figure}

Figure~\ref{MF_graph} shows good agreement with the MF approximation, given in Eq.~(\ref{eq:MF}) in the low-density regime. As follows from this MF approximation,
there is no dependence on the field $G$ at all. Above $\rho \approx 0.2$, correlation start playing a substantial role and the observed current is lower than the MF prediction. As $G$ increases, the rate to move in the direction transverse to the field is lower, thus creating more constrained dynamics, leading to stronger deviation from MF behavior. We also observe that, although our MF theory predicts $J>0$ for any density $\rho<1$, in our simulations the current stops at some finite density, which is dependent on the field $ G $. Specifically, for $G=0$, jamming occurs at $\rho=0.83$ and for $ G=\infty$, it occurs at $\rho=0.78$. This jamming at finite density has been observed in the $E=\infty$ limit of the E-model~\cite{Sellitto_ASEP_KCM_PRL}, which coincides with our $G=0$ limit.

Note that these values of the jamming density are significantly lower than what we would expect in the Kob-Andersen model at such system sizes without driving. Although in the thermodynamic limit the undriven model jams only at $\rho=1$~\cite{TBF}, finite size effects lead to jamming at $\rho_c = 1 - \frac{\lambda}{\log L}$, where $\lambda$ may be approximated by $\lambda=0.25$~\cite{Teomy_JCP}. Thus for our $L=400$ system size, this would lead to $\rho_c=0.96$.

\subsection{Current as a function of $G$}

\begin{figure}[H]
\includegraphics[width=\columnwidth]{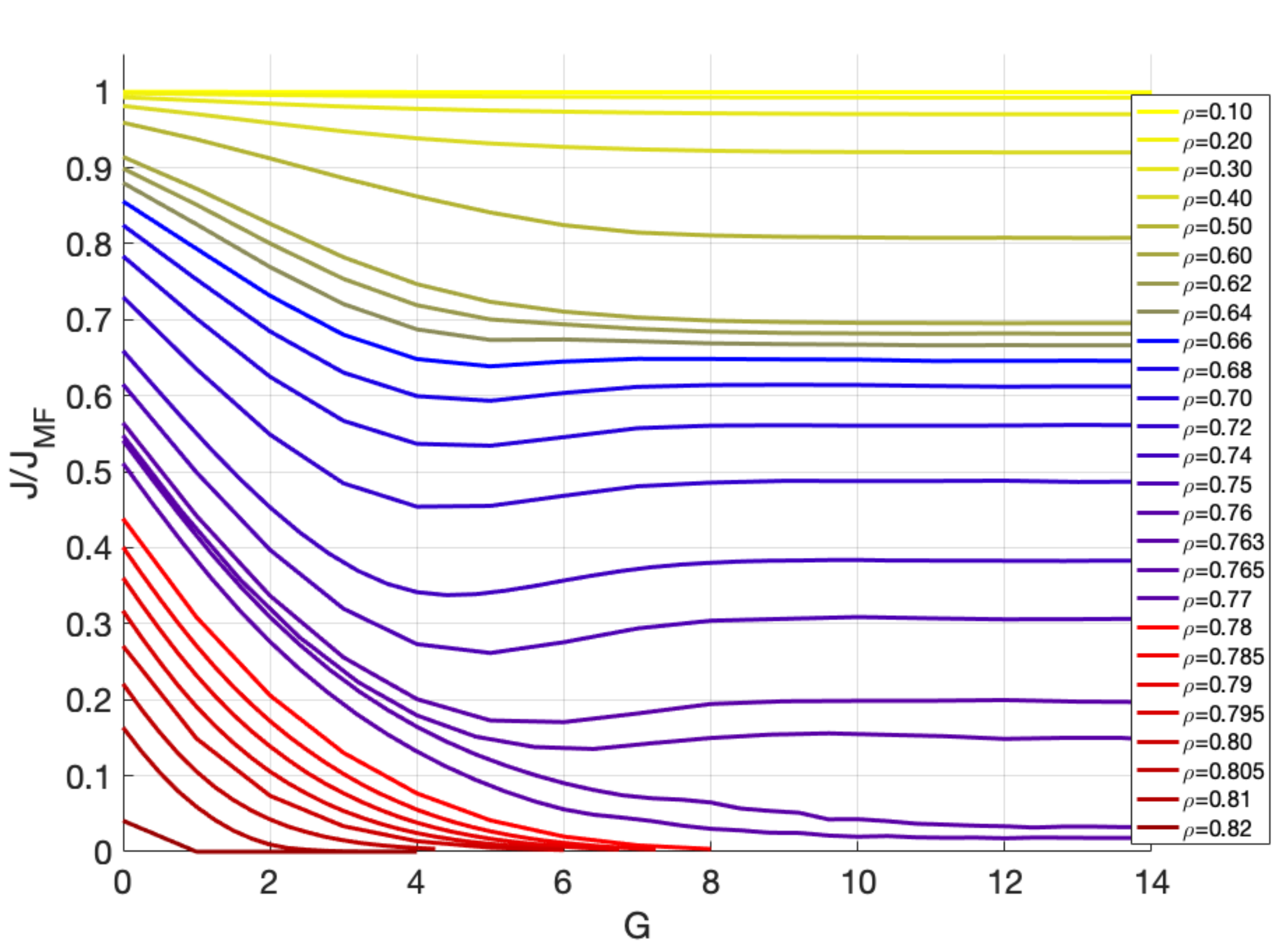}
\caption{Normalized Current $ J/J_{MF} $ as a function of the external field G. For lower densities, the deviation from the mean-field current is small. This deviation increases as we increase G. As the density grows, the current becomes non-monotonic as a function of $G$ for $0.66 \leqslant \rho \leqslant 0.76$.}
\label{fig:J_vs_G}
\end{figure}

We simulated the steady state current behavior as a function of the external field $G$. The results of the simulation for varying $G$ values can be seen in Fig.~\ref{fig:J_vs_G}. For each density, we normalize the current by the MF prediction according to Eq.~(\ref{eq:MF}). As we have already seen, the substantial deviation from the MF result, and the significant dependence on $G$, both take place at higher densities. This can be understood by the fact that the rate to move sideways controls the probability to get stuck in the way. Whenever the rate to move sideways is large ($G$ is low), there is a higher probability that a particle will change it's trajectory and hence will be less correlated, causing the current to increase at lower values of $G$.

Remarkably, in a finite region of densities, $0.66 \leqslant \rho \leqslant 0.76$, the dependence on $G$ becomes non-monotonic and there is a region where the current slightly increases as we increase $G$ until the current reaches its asymptotic value at infinitely large values of $G$. 

\subsection{Jamming phase diagram}

\begin{figure}[h]
\includegraphics[width=\columnwidth]{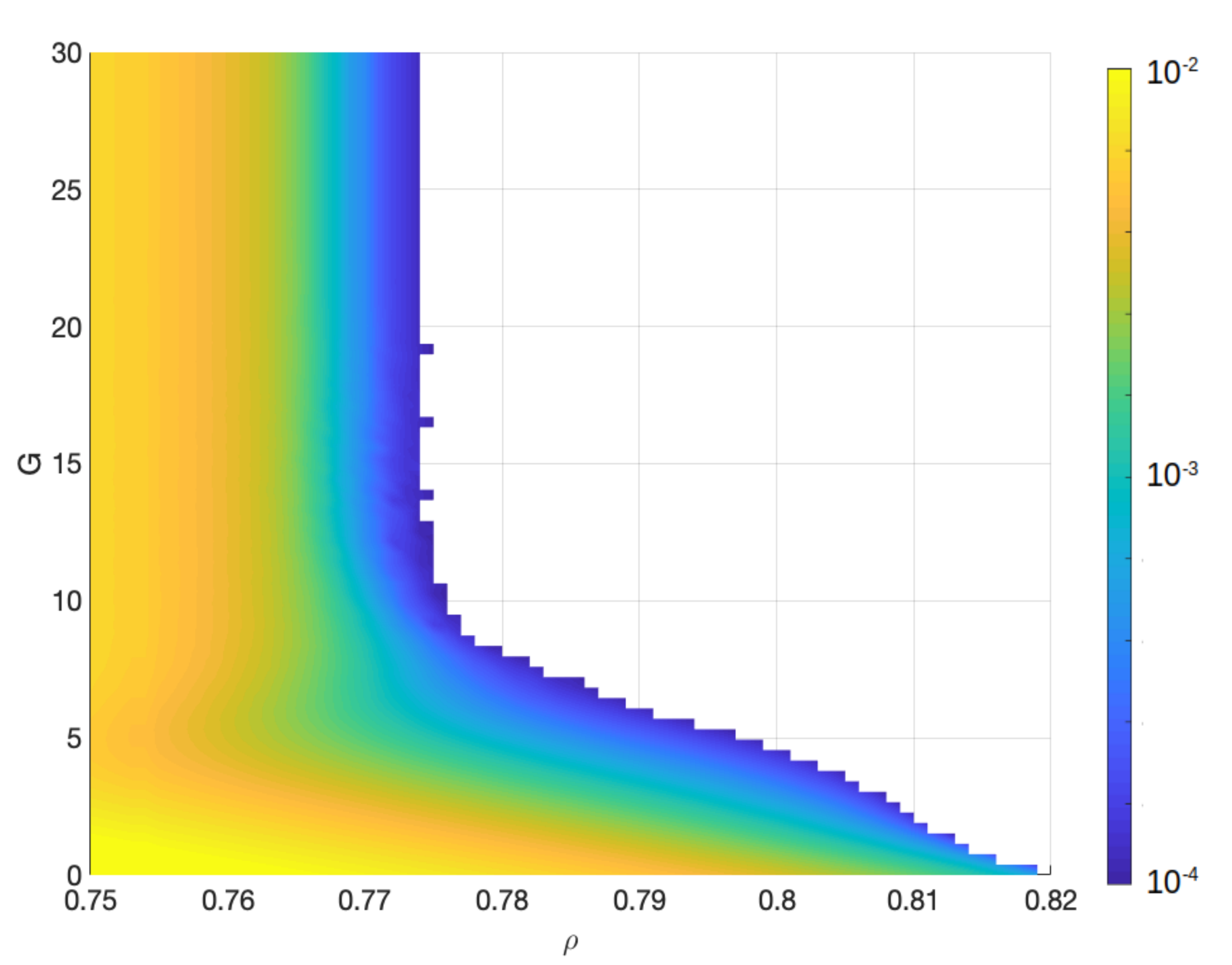}
\caption{Phase diagram in the density-field plane showing the jamming transition. The colored region is the flowing state with steady state current values depicted as the colorbar, and the white space is the jammed state where current vanishes.}
\label{fig:phase_diagram}
\end{figure}

According to out MF prediction, the current asymptotically vanishes only as the density approaches unity. However, from the numerical observations, one can see that there is some field-dependent critical density $\rho_c(G)$, at which the particles are arrested and the system gets jammed. Specifically, Fig.~\ref{MF_graph} shows that the current goes to zero at different densities for $G=0$ and $G=\infty$. This implies that such jamming effect depends on the field intensity $G$. In order to investigate the transition from free flow to a jammed state, we obtained the phase diagram shown in Fig.~\ref{fig:phase_diagram}, which describes the numerical results of the current as a function of the density and field.

As we can see from the phase diagram, the transition to the jammed state happens at the density range  $0.78<\rho_c(G)<0.83$. This critical density was observed in the past for the $ E=\infty$ limit of the E-Model \cite{Turci_Pitard_Driving_KCM}, which is equivalent to $G=0$ in our G-Model. However, we have achieved the extension to higher values of $G.$ We find that as the density increases, a smaller field is required to get the jammed phase.

\subsection{Correlation analysis}

\begin{figure}[b]
\includegraphics[width=\columnwidth]{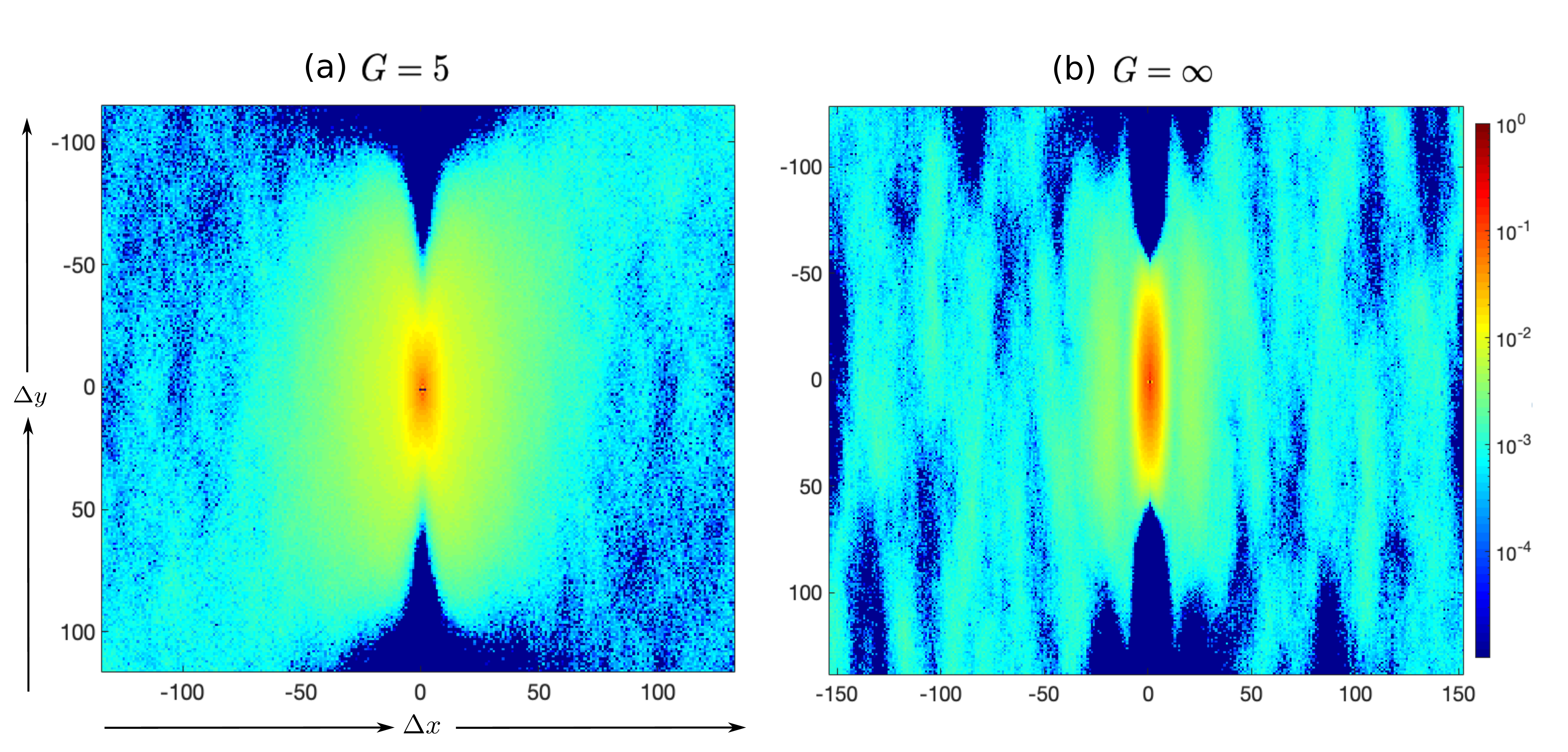}
\caption{Two-point density correlation function in the steady state with a system size $L=400$ having density $ \rho=0.07.$ Left and right panels are for $G=5$ and $G=\infty$, respectively.}
\label{fig:Cor_map}
\end{figure}

We have performed correlation analysis by defining the two-point density correlation function as,
\begin{equation}
C_{2}(\Delta x,\Delta y)=\frac{\left\langle \eta_{i,j}\eta_{i+\Delta x, j+\Delta y}\right\rangle -\rho^{2}}{\rho-\rho^{2}} .
\end{equation}
As can be seen from Fig.~\ref{fig:Cor_map}, the correlations are mostly longitudinal as $G$ becomes stronger. This is because increasing $G$ is equivalent to decreasing the probability to move in the transverse direction, which corresponds to the dominant motion being longitudinal.

\subsection{Analysis of microscopic structure}

\begin{figure}[t]
\includegraphics[width=\columnwidth]{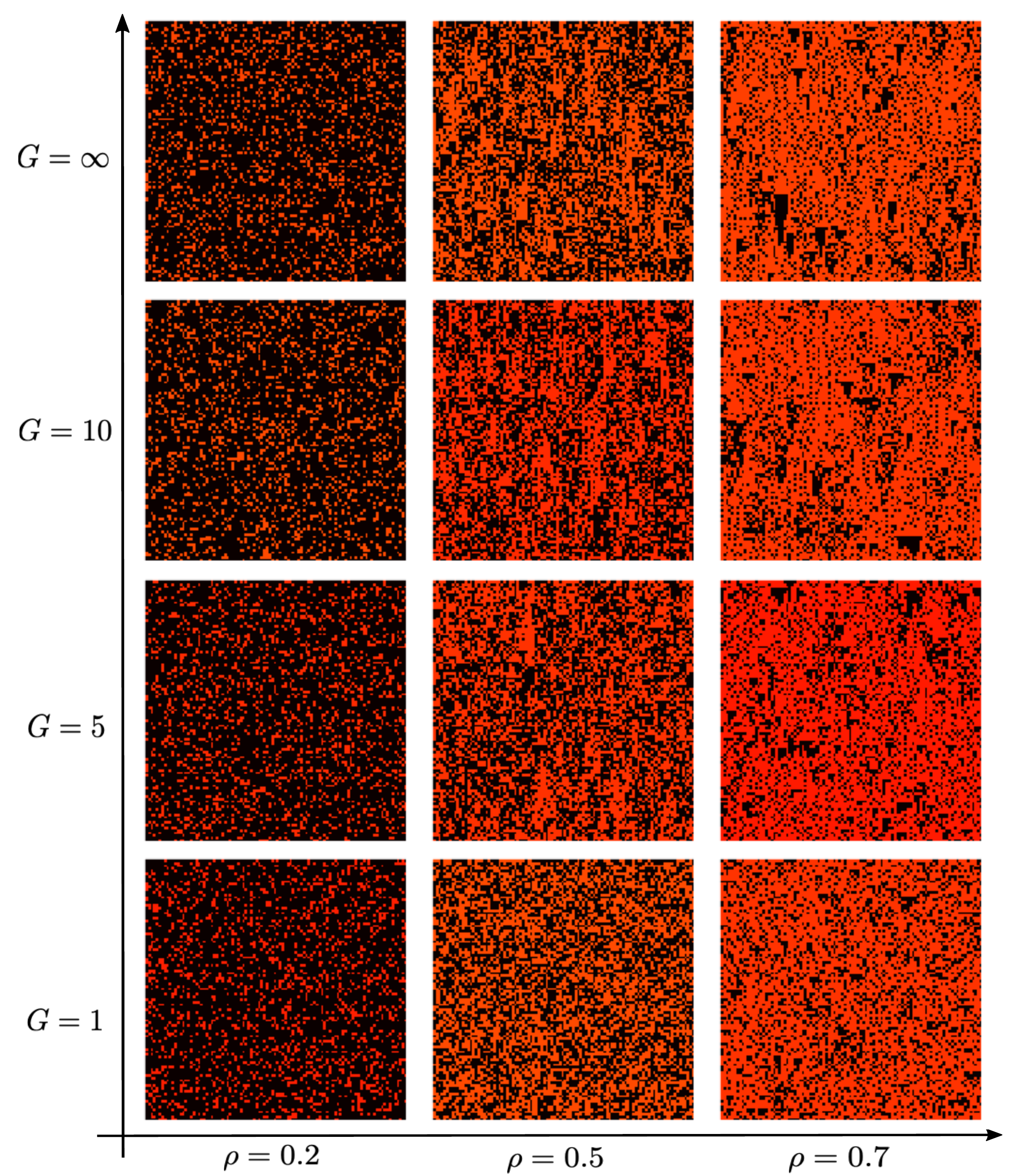}
\caption{Structures of particles (red) and holes (black) in steady state for different values of field and density.}
\label{fig:rho_vs_G_structures}
\end{figure}

\begin{figure}[t]
\begin{minipage}[t]{0.45\columnwidth}%
\begin{center}
\subfloat[Low $G$\label{fig:Struct_low_G}]{\begin{centering}
\includegraphics[width=3.5cm]{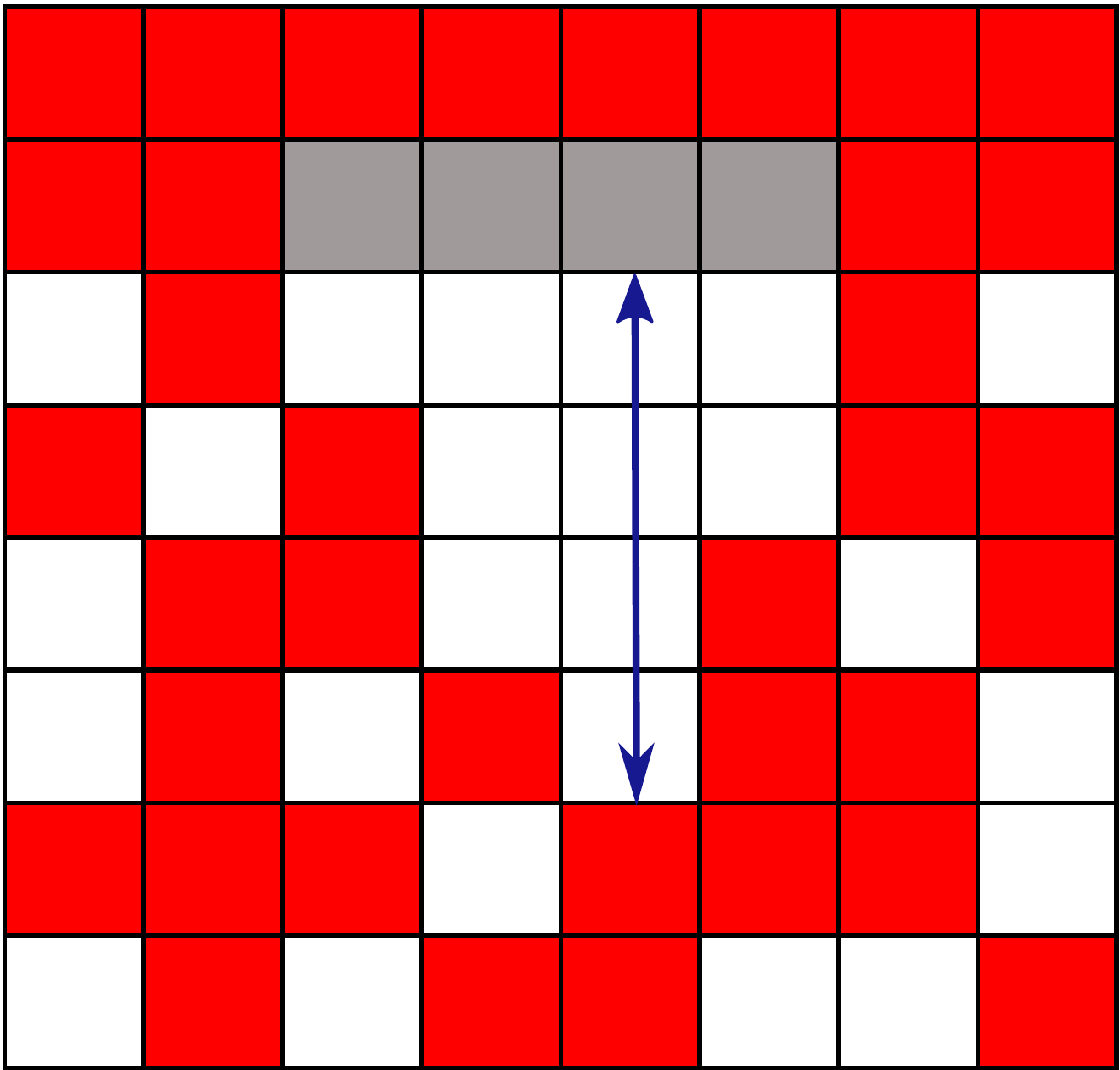}
\par\end{centering}
	}
\par\end{center}%
\end{minipage}\hfill{}%
\begin{minipage}[t]{0.45\columnwidth}%
\begin{center}
\subfloat[High $G$\label{fig:Struct_high_G}]{\begin{centering}
\includegraphics[width=3.5cm]{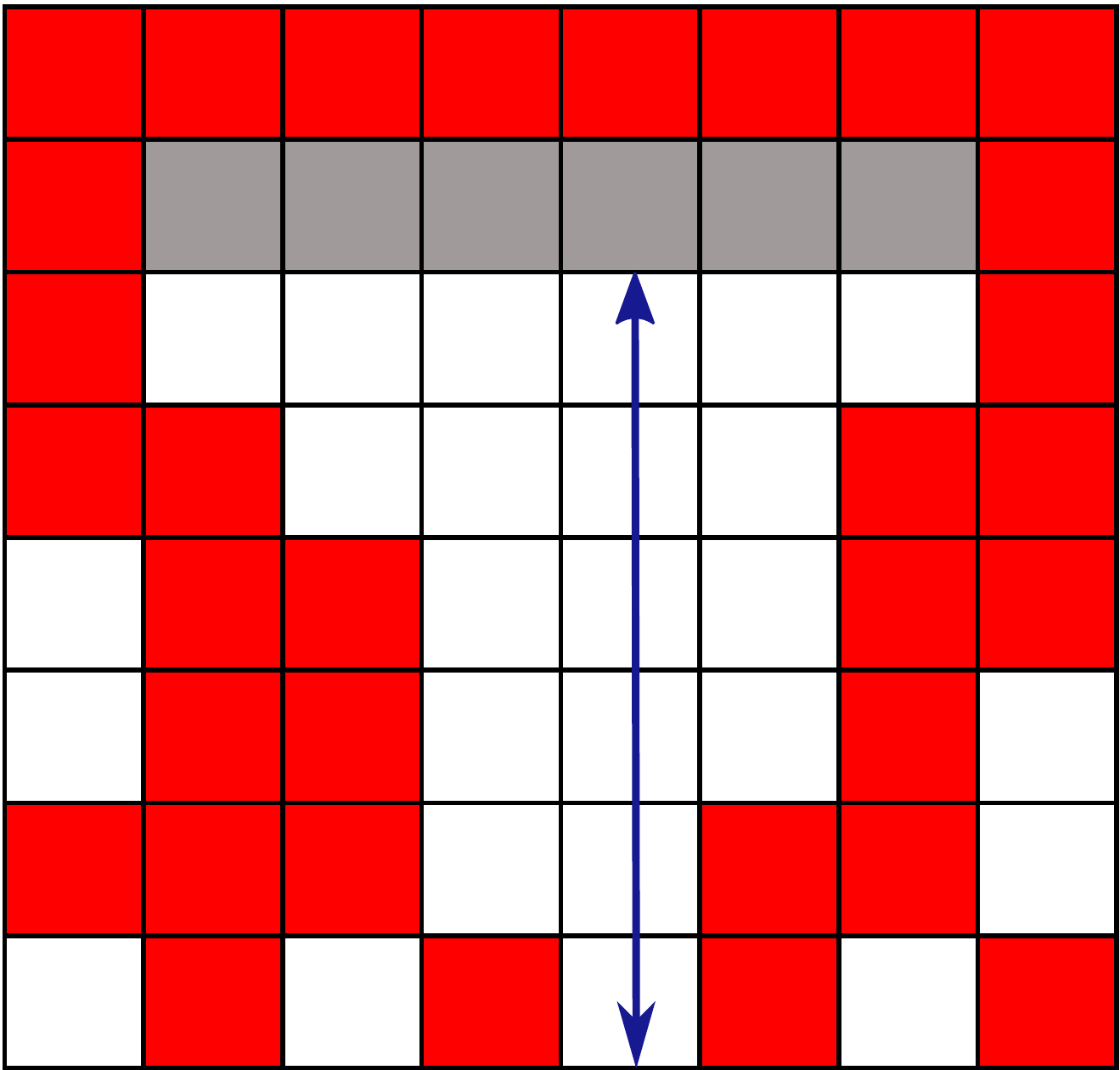}
\par\end{centering}
	}
\par\end{center}%
\end{minipage}
\caption{(a) Typical structure of the hole clusters for small values of $G$. The smaller shear band (gray), together with a higher moving rate in the transverse direction, result in a smaller height of the hole cluster. (b) Typical structure of the hole clusters for high values of $G$. Lower transverse rate results in a higher height to fill in the vacancies below the wider shear band.}
\end{figure}

One of the prominent differences between various regimes of $G$ values are the internal structures that are formed in the steady state. As the density increases, the homogeneously distributed particles start to create clusters of particles and clusters of holes, as can be observed from Fig.~\ref{fig:rho_vs_G_structures}. At the high densities, the typical shapes of the microscopic structures of vacancies  are illustrated in Fig.~\ref{fig:Struct_high_G}. The formation of these hole clusters starts with the formation of shear bands~\cite{Turci_Pitard_Driving_KCM} which are perpendicular to the applied field. These shear bands basically act as barriers to the flow, below which there are areas of vacancies, which go around these barriers and, due to transverse motion, gradually fill in these areas of vacancies.

As can be seen from Fig.~\ref{fig:Struct_low_G}, for small values of $G$, the height of the vacancies region is typically small, as there is a high transverse motion, and the particles fill the open space below the shear bands relatively quickly, as opposed to the typical structure of the hole cluster for high values of $G$, as illustrated in Fig.~\ref{fig:Struct_high_G}. As we further observe, these hole clusters are not persistent and they dynamically form and get destructed in the steady state. The evolution of the configuration at the steady state and the appearance of the hole clusters can be seen in Fig.~\ref{fig:rho_vs_G_structures}. This effect starts taking place at high densities. For $G=\infty$, there is no motion at all in the transverse direction and the only motion is in the direction of the field. In this case, it can be seen that the hole clusters are the longest, as depicted in Fig.~\ref{fig:Struct_high_G}.

\section{Flow through an orifice}\label{sec:orifice}

\begin{figure}[b]
\includegraphics[width=0.5\columnwidth]{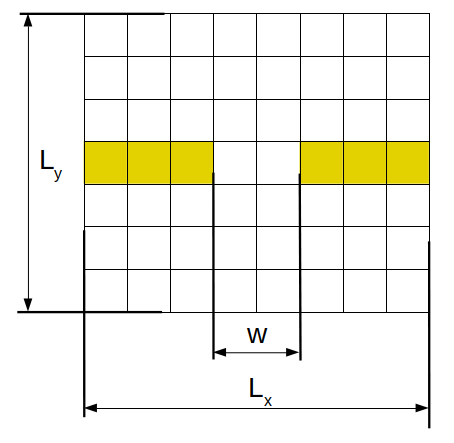}
\caption{Schematic diagram of a narrow opening of width $w$ within a horizontal wall of immobile particles (yellow).}
\label{orifice_geometry}
\end{figure}

In this section, we study our model with confinement that mimics the behavior of discharging granular materials through an opening. The confinement reveals new phenomena, such as density heterogeneity and clogging of the system near the opening. In addition, we would like to investigate the confined flow and to reveal its connection to the bulk current, discussed in Sec.~\ref{sec:bulk_current}.

\begin{figure}[t]
\includegraphics[width=\columnwidth]{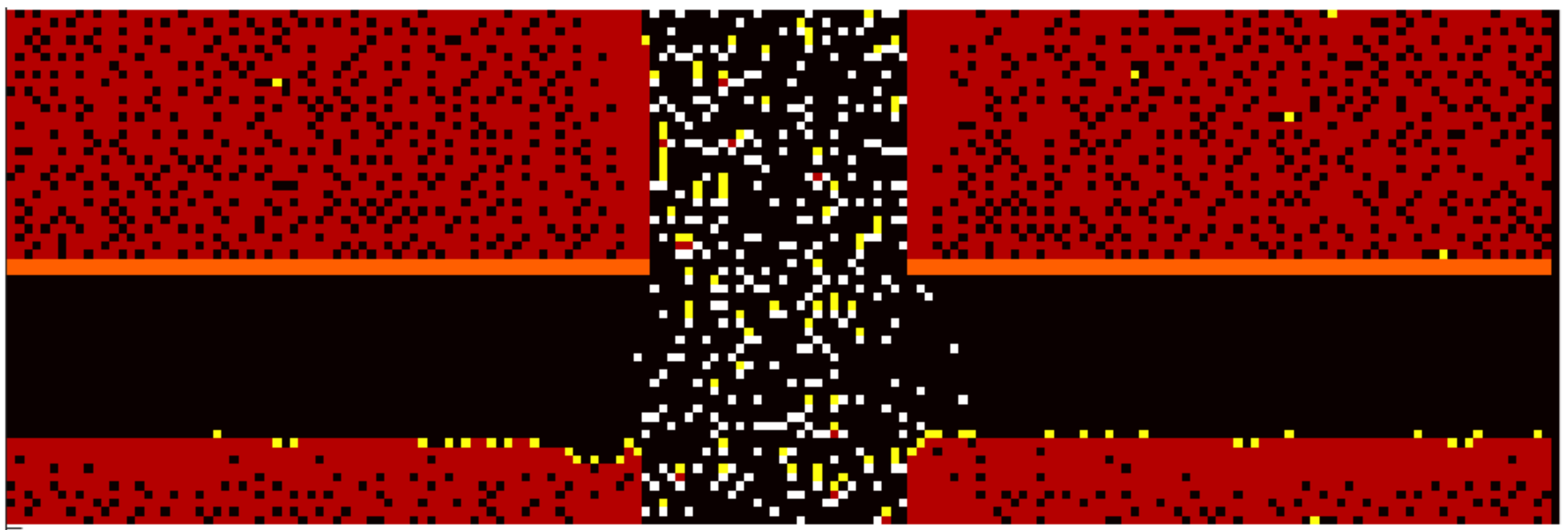}
\caption{Snapshot of particle configuration in a lattice of dimension $60\times180$ with an orifice of width $w=30$, at field $G=1$, and density $\rho=0.5$. The system is in steady state that was reached after $10^6 $ units of time. Red particles are stuck (non-movable in any direction), white particle can move in at least one direction, black sites are empty.}
\label{steady_state_formation}
\end{figure}

We define a confined geometry imitating the discharge of grains through an orifice, as shown in Fig.~\ref{orifice_geometry}. Also in this geometry, we employ periodic boundary condition in both directions, whereas, at the center of the lattice we place a rigid horizontal barrier, with a narrow opening of width $w$. The obstacle is achieved by artificially placing one row of occupied sites in the initial condition and not moving them during the simulation. Due to the periodic boundary conditions, this geometry is equivalent to infinitely many openings with distances $L_x-w$ in the transverse direction, and infinitely many such barriers at a distance $L_y$ in the longitudinal direction. In order to eliminate the finite size effect, we maintained a considerable value of the ratio $L_x/w$.

The simulation begins with randomly and uniformly distributed particles throughout the lattice. Due to the presence of the barrier, there is sedimentation of the particles towards the barrier, whereas in front of the opening there is flow. As a result, heterogeneity of the density develops. For $G<\infty$, due to lateral motion, particles rearrange, such that the density above the barrier increases, and by mass conservation, the density above the opening decreases, as shown in Fig.~\ref{steady_state_formation}. We also observe that in the steady state, the only contributing part to the current is the flow above the orifice, denoted by the yellow particles in Fig.~\ref{steady_state_formation}. This is because the system is jammed in other regions of the lattice, as shown by the light blue particles.

We are interested in measuring the \textit{total current} of the particles through the opening, namely the number of particles flowing through the opening per unit time as given by,
\begin{equation}
J_{\textrm{tot}}=\frac{\left\langle N \right\rangle}{t L_y} , \label{eq:vol_flow_def}
\end{equation}
where $ \left\langle N \right\rangle $ is the average number of particles that moved along the direction of the field during the time interval $t$ anywhere in the lattice, and $L_y$ is the lattice height (see Fig.~\ref{orifice_geometry}). Using this definition, we can now estimate the total current through the orifice using the bulk current, which we studied in Sec.~\ref{sec:bulk_current}. Recalling that the bulk current is defined as the number of particles flowing in the system per unit time and per lattice size, we can formally define the bulk current as, $J_{\textrm{bulk}}=\left\langle N \right\rangle/(t L_x L_y).$ In order to be consistent with the definition of the total current in Eq. (\ref{eq:vol_flow_def}), our crudest approximation will be as follows,
\begin{equation}
J_{\textrm{tot}}=J_{\textrm{bulk}}(\rho,G)w \label{vol_flow_analyt_bulk}
\end{equation}

\begin{figure}[t]
\includegraphics[width=\columnwidth]{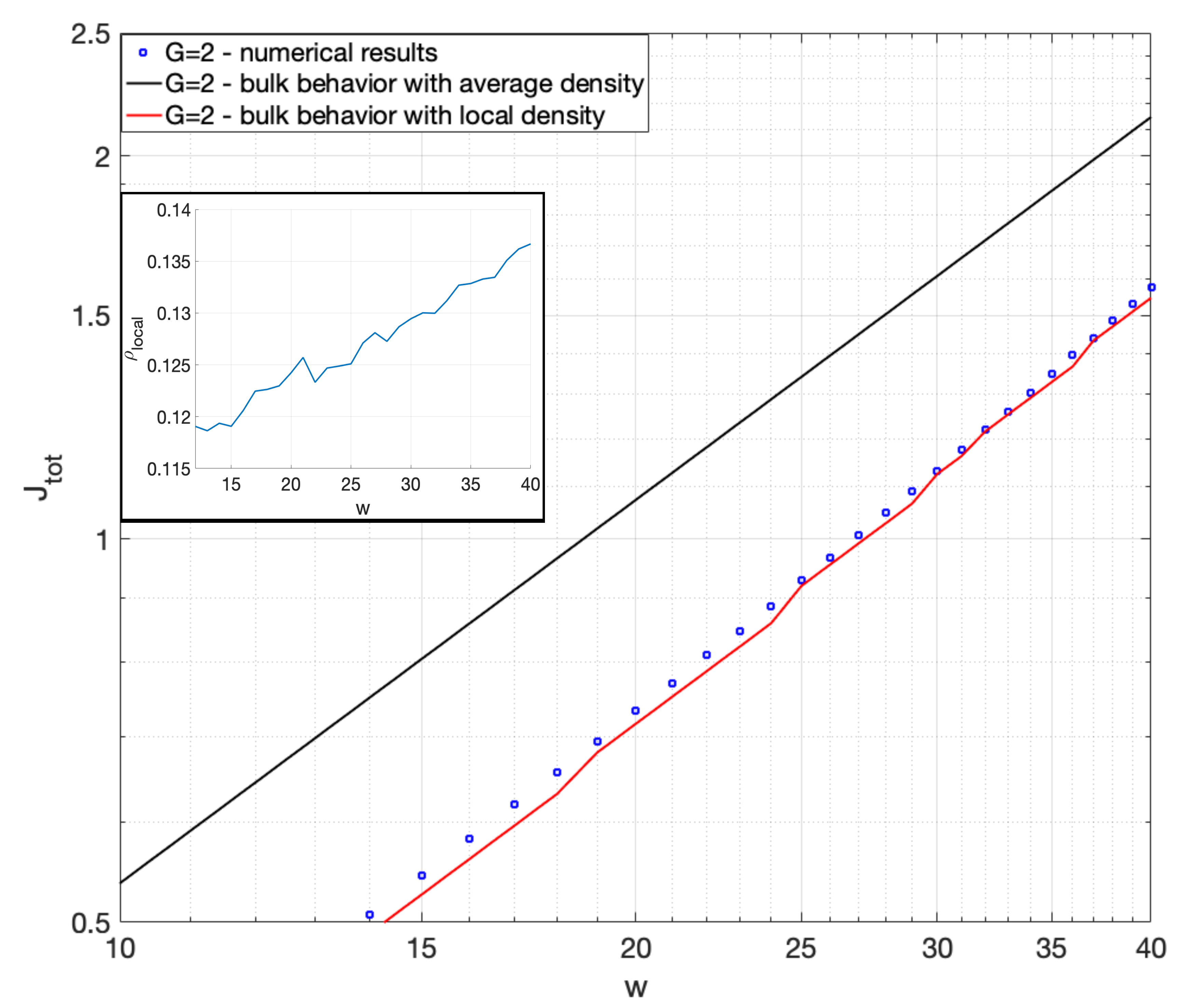}
\caption{Total current $J_{tot}$ vs. orifice width $w$ at bulk density $\rho=0.5$ with $G=2$ in a $(60\times180)$ lattice. Analytical expression, Eq.~(\ref{vol_flow_analyt_bulk}) using bulk density (black) does not match with simulation (blue points), whereas the analytical expression, Eq.~(\ref{vol_flow_analyt_local}) using local density (red) shows a very good agreement with simulation. Inset: local density above the opening vs. opening width. }
\label{fig:J_vs_W_G_2}
\end{figure}

Figure~\ref{fig:J_vs_W_G_2} shows results of numerical simulation, where we calculated the total current through the orifice, in comparison to the crude approximation given in Eq.~(\ref{vol_flow_analyt_bulk}). As we can see from the plot, there is a considerable discrepancy between the numerical results (blue points) and this calculation (red line). To explain this, we recall that the density in the confined system is heterogeneous, and in particular, the density above the orifice is lower than the average density in the system. Hence we consider the local density measured above the orifice, as this is the only region that contributes to the current in the steady state. This corrects our theoretical prediction to,
\begin{equation}
J_{\textrm{tot}}=J_{\textrm{bulk}}({\rho_\textrm{local}},G)w . \label{vol_flow_analyt_local}
\end{equation}

\begin{figure}[t]
\centering
\includegraphics[width=\columnwidth]{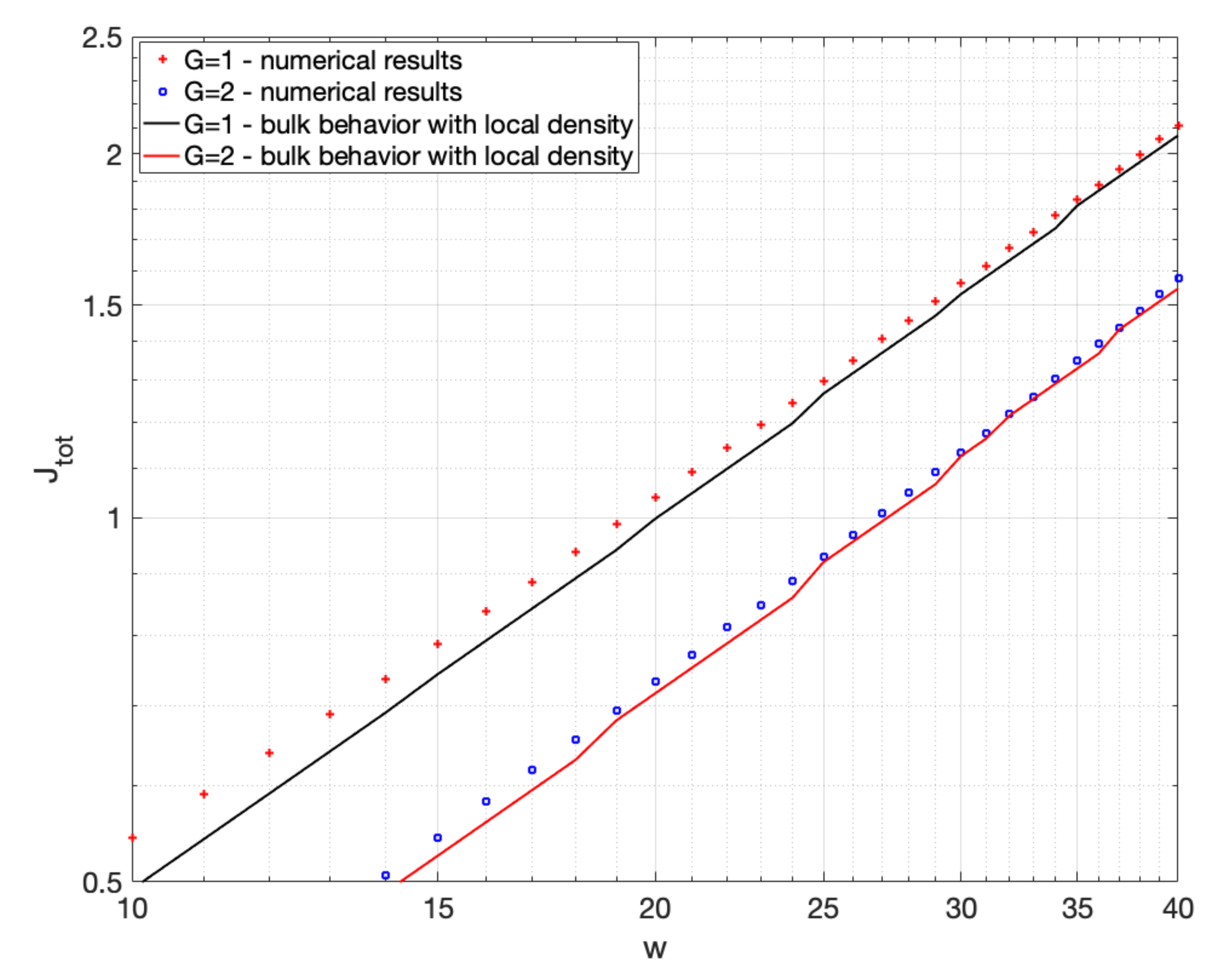}
\caption{Total current $J_{tot}$ vs. orifice width $w$ for $G=1$ (black) and $G=2$ (red) at bulk density $\rho=0.5.$ Data points from numerical simulation and solid lines from analytical expression, Eq.~(\ref{vol_flow_analyt_local}), using the local density measured in the simulations. \label{numerical_bulk_dif_g}}
\end{figure}

In Sec.~\ref{sec:bulk_current}, we have seen that in non-confined systems, the bulk current is higher for lower densities. Remarkably, the analytical expression~(\ref{vol_flow_analyt_local}) that takes into account the reduced local density above the opening for the total current coincides with the numerical results, as seen in Fig.~\ref{fig:J_vs_W_G_2}. Thus, using the behavior studied for the bulk system, we can predict the dynamics in confined systems. We can perform a similar comparison also for different values of $G.$ As the field increases, the transverse motion becomes less probable due to a lower rate to move in that direction. This leads to a longer time needed to reach the steady state in confined systems. In Fig.~\ref{numerical_bulk_dif_g} we show the numerical results for two different values of $G$ with remarkable agreement with the same analytical prediction, described by Eq.~(\ref{vol_flow_analyt_local}). As $w$ becomes smaller, the current decreases linearly with $ w $, as Eq.~(\ref{vol_flow_analyt_local}) predicts. For very small values of $ w $, we obtain clogging of the particles near the opening and the overall current ceases for some realizations. The numerical results shown in Fig.~\ref{fig:J_vs_W_G_2} and \ref{numerical_bulk_dif_g} are obtained by averaging over the current from many realization that weren't clogged. This clogging state becomes more and more probable as $ w $ becomes smaller, and it would be interesting to study it further.

\section{Conclusions}\label{sec:conclusions}

We studied the behavior of granular materials subjected to a gravitational field in a confined geometry. To that end, we used the kinetically-constrained Kob-Anderson model, which implements simple dynamics of a lattice gas, with no interactions between neighboring particles, but with kinetic rules that depend on the occupancy of neighboring sites. Such a description allows us to employ efficient numerical simulations. In order to simulate the gravitational field, we modified the rates of the moves in the direction perpendicular to the field, controlled by the  parameter $ G $, and did not allow moves against the direction of the field. We first studied the bulk behavior of the model, eliminating confinement by creating a large system with periodic boundary conditions. We built a MF theory and performed numerical simulations in order to understand the range of validity of the MF theory and the role of correlations. For that purpose, we constructed an efficient rejection-free algorithm, in which we skip all the unsuccessful moves. We observed that at the range of densities $ 0.66 \leqslant \rho \leqslant 0.76 $ the current exhibits non-monotonic behavior as a function of the driving field $ G $. From the microscopic point of view we see that at the high density regime, close to jamming, the spatial distribution of the particles is very heterogeneous, namely there are spatial structures of holes that are dynamically formed and destructed as a result of the development of shear bands. 

To study discharge through a narrow opening, we defined a confined geometry in our 2D lattice by introducing a horizontal barrier with an opening of some size that we could change. We conducted numerical simulations, measuring the total current through the opening, and revealed two phenomena - clogging close to the opening and highly reduced density of the flowing particles. We measured the spatial distribution of the particles in the steady state with confinement, measuring only the current above the opening, as this is the only contributing part to the current in the whole system. We were able to analytically predict the current as a function of the opening width by using the bulk current  corresponding to the reduced local density, as measured from numerical simulations with confinement. We obtained excellent agreement between the total current through the opening from numerical simulation and the analytical calculation based on the current measured in the bulk case, corresponding to the reduced local density above the opening.

One of our most remarkable observations for bulk behavior is the non-monotonicity of the current as a function of field intensity at intermediate densities. Further structural analysis could identify the reason for this behavior. 

Similar studies of discharge through an opening and related phenomena such as spatial density heterogeneity and bottleneck clogging have already been seen in granular matter, whether in experiments or simulations. Our model reveals qualitatively the same phenomena, whereas thanks to its simplicity the simulations are fast and can be done on very large systems. The model can also be extended to three dimensions where, instead of looking on the square lattice, one may consider the cubic lattice. This case may perhaps be more suitable for comparison with actual granular experiments.

\section{Acknowledgments}
We thank David Gomez, Erdal Oguz, Daniel Vilyatser, Eyal Rubin, Nimrod Segall,  Hadas Shem-Tov, and Eial Teomy for helpful discussions. This research was supported by the Israel Science Foundation Grant No. 968/16 and by the Prof. A. Pazy Research Foundation.

\end{document}